# Electronic and vibronic properties of a discotic liquid-crystal and its charge transfer complex

Lucas A. Haverkate[1], Mohamed Zbiri[2,a], Mark R. Johnson[2], Elizabeth Carter[3], Arek Kotlewski[4], S. Picken[4], Fokko M. Mulder[1], Gordon J. Kearley[5]

[1]*Reactor Institute Delft, Faculty of Applied Sciences, Delft University of Technology, Mekelweg 15, 2629JB Delft, The Netherlands*

[2]*Institut Laue Langevin, 38042 Grenoble Cedex 9, France*

[3]*Vibrational Spectroscopy Facility, School of Chemistry, The University of Sydney NSW 2006, Australia*

[4]*ChemE-NSM, Faculty of Chemistry, Delft University of Technology, 2628BL/136 Delft, The Netherlands*

[5]*Bragg institute, Australian Nuclear Science and Technology Organisation, Menai, NSW 2234, Australia*

[a] *zbiri@ill.fr*

Discotic liquid crystalline (DLC) charge transfer (CT) complexes combine visible light absorption and rapid charge transfer characteristics, being favorable properties for photovoltaic (PV) applications. We present a detailed study of the electronic and vibrational properties of the prototypic 1:1 mixture of discotic 2,3,6,7,10,11-hexakishexyloxytriphenylene (HAT6) and 2,4,7-trinitro-9-fluorenone (TNF). It is shown that intermolecular charge transfer occurs in the ground state of the complex: a charge delocalization of about $10^{-2}$ electron from the HAT6 core to TNF is deduced from both Raman and our previous NMR measurements (Ref. 32), implying the presence of permanent dipoles at the donor-acceptor (D-A) interface. A combined analysis of density functional theory calculations, resonant Raman and UV-VIS absorption measurements indicate that fast relaxation occurs in the UV region due to intramolecular vibronic coupling of HAT6 quinoidal modes with lower lying electronic states. Relatively slower relaxation in the visible region the excited CT-band of the complex is also indicated, which likely involves motions of the TNF nitro groups. The fast quinoidal relaxation process in the hot UV band of HAT6 relates to pseudo-Jahn-Teller interactions in a single benzene unit, suggesting that the underlying vibronic coupling mechanism can be generic for polyaromatic hydrocarbons. Both the presence of ground state CT dipoles and relatively slow relaxation processes in the excited CT band can be relevant concerning the design of DLC based organic PV systems.

# I. INTRODUCTION

Discotic liquid crystals (DLCs) are considered as a promising class of organic materials for photovoltaic (PV) and other electronic applications.[1-7] These disk-like molecules form stable columns due to the π-π orbital overlap of their aromatic cores, while thermal fluctuations of their side chains give rise to the liquid-like dynamic disorder.[8] DLCs combine advantageous materials properties, including visible light absorption, long-range self-assembly, self-healing mechanisms, high charge-carrier mobilities along the column axis and a tunable alignment of the columns. Like conjugated polymers, DLCs offer the potential of low-cost, easily processed and flexible solar cells.[4,9] But, a major challenge to application is to achieve a morphology that enables a bulk-heterojunction (BHJ) PV device architecture.[6,10] This topic has attracted considerable interest over the past decade, with the goal of obtaining an interpenetrating network of electron and hole conducting molecular columnar wires enabling D-A phase separation on the nanoscale.[1,3,11,12]

Another crucial issue for organic PV application is the strong electron-phonon coupling inherent to molecular systems, which limits the efficiency of charge separation.[10] Upon photoexcitation, strongly bound exciton states are formed that first need to dissociate before charge transport to the electrodes can occur. Dissociation takes place at the D-A interface, where intermediate excited charge-transfer (CT) states are formed with the hole on the donor and electron at the acceptor molecule. At present time there is no clear picture of the final charge separation process, i.e. the dissociation of the excited CT state into a free electron and hole. The lowest CT state corresponds to a Coulombically bound electron-hole excited pair, with a binding energy of typically several hundreds of meV.[13] Various scenarios have been proposed to explain separation of charges from this lowest state, including the possible presence of dipoles at the D-A interface.[10,14] Other studies indicate that charge separation is mediated by the higher lying vibronic states of the excited CT manifold.[15,16] In such a process, charge carriers undergo a few ultrafast hops via an activationless pathway, allowing their separation before the thermal relaxation and towards the occurrence of localized excitonic levels. In this respect, fundamental knowledge of the electronic and vibrational properties of the excited state levels and relaxation pathways is a key topic in further understanding and improving the working principles of OPVs.[15]

This paper investigates the possible presence of permanent dipoles at the D-A interface of HAT6-TNF compound. Further, a first step is presented to characterize the influence of molecular vibrations on the charge carrier relaxation in self-assembled DLCs. For self-assembled aggregates, such as DLCs and DLC ground state CT complexes, the characterization of photo-induced electron transfer and relaxation processes is at an early stage.[17] The addition of electron acceptors has been shown to increase the conductivity of DLCs.[18-20] On the other hand, it has been proposed that recombination processes limit the hole photocurrent in such compounds.[21] Charge carriers in ground state CT compounds are supposed to be trapped and readily annihilated through rapid, phonon-assisted relaxation and recombination processes.[3,21,22]

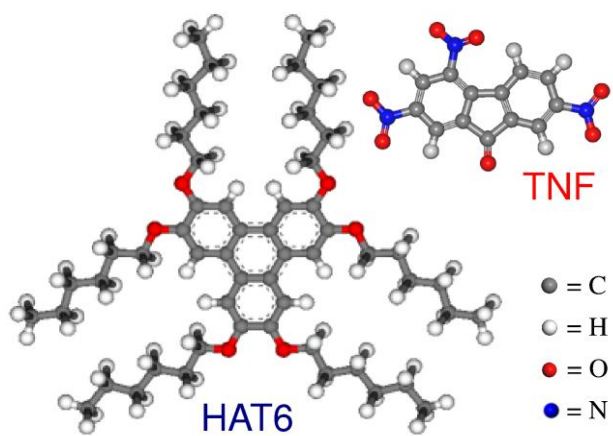

FIG. 1. Illustration of HAT6 (hexakis(n-hexyloxy)triphenylene) and TNF (2,4,7-trinitro-9-fluorenone) molecules.

The prototypical discotic electron donating discoid HAT6 and its 1:1 mixture with electron acceptor TNF (Fig. 1) is chosen as a model system. HAT6-TNF forms a ground state CT compound exhibiting a stable columnar phase from below room temperature to 237 $^{0}$C.[23] The high symmetry and moderate molecular size of discogens such as HAT6 makes these systems attractive for exploring the effects of increasing molecular complexity, by comparing their photophysical properties with those of the fundamental building block, benzene, and large polyaromatic hydrocarbons.[24-26]

We start with an investigation of the electronic properties of HAT6-TNF. For discotic liquid crystalline complexes, formed by donor and acceptor molecules, it is generally accepted that the intermolecular charge transfer is of an excited state nature.[27,28] Indeed, mixtures of the electron donating discoids with non-discogenic electron acceptors exhibit absorption bands in the visible region due to excited state charge transfer.[29-31] However, recently we found indications of weak electron transfer occurring in the ground-state of HAT6-TNF.[32] This implies the presence of permanent dipoles between the donor (HAT6) and acceptor (TNF) molecules. Here, we find strong support for these indications from a combination of our previous NMR results[32] with the present Raman spectroscopy measurements. In addition, we characterize the electronic transitions involved in the excited CT-band of HAT6-TNF by combining UV-VIS absorption and resonant Raman spectroscopy. Subsequently, the UV-VIS and Raman measurements are combined with density functional theory (DFT) calculations, to identify the vibrational modes that assist charge-carrier relaxation in the "hot" band of HAT6 and in the excited CT-band of HAT6-TNF.

## II. MATERIALS AND METHODS

### A. Sample preparation.

Isotopically normal 2,3,6,7,10,11-hexakishexyloxytriphenylene (HAT6) and its side-chain deuterated analog, HAT6D, were prepared by the synthesis methods described earlier[23,33]. The charge transfer compounds were obtained by mixing HAT6 (or HAT6D) with 2,4,7-trinitro-9-fluorenone (TNF) in a 1:1 molar proportion in dichloromethane.[23] The mixture was subsequently evaporated to dryness at room temperature. To remove any traces of solvent and to ensure the correct phase behavior, the resulting composite was heated to the isotropization temperature, T=237 $^0$C, then cooled slowly. By using a deuterated analog for TNF as well (TNFd, with all hydrogens deuterated), four different analogues were obtained: HAT6-TNF, HAT6D-TNF, HAT6-TNFd and HAT6D-TNFd. The degree of deuteration of HAT6D and TNFd was about 98 atom%.

### B. Absorption spectroscopy.

Optical absorption at room temperature was measured using a Perkin-Elmer Lambda 900 spectrometer equipped with an integrating sphere. The optical density was measured and the attenuation $F$a (fraction of incident photons that is absorbed by the sample) was obtained by correction for reflection losses.

## C. Raman spectroscopy.

Spectra at wavelengths of 1064, 785, 633, 514 and 488 *nm* were collected at the Vibrational Spectroscopy Facility (School of Chemistry, The University of Sydney (USYD)). The off-resonance spectra at 1064 *nm* were recorded with a Bruker FT-Raman (MulitRAM) spectrometer using a ×100/1.25 *NA* objective, with the laser power at the sample spot between 50 and 200 *mW* depending on the sample. The spectra at 785, 633, 514 and 488 *nm* were obtained with a Renishaw Raman InVia Reflex Microscope (Renishaw plc., Wotton–under–Edge, UK), using a ×50/0.75 *NA* objective and a laser power of 0.1-1.0 *mW*. The details of this spectrometer have been described elsewhere.[34] A Renishaw Raman InVia Reflex Microscope (Renishaw plc., Wotton–under–Edge, UK) spectrometer located at the Analytical Centre, The University of New South Wales (UNSW) was used to collect spectra at an excitation wavelength of 325 *nm*, the samples being measured with a ×40 objective and the laser power at the sample being estimated as between 0.8-1.0 *mW*. Spectra were obtained from different positions on a selected sample region, the number of selected spots varying between 10 and 50 depending on the sample and wavelength, these spectra were then averaged. The accumulation and exposure times were typically 10-50 and 10 *s*, respectively.

## D. Raman simulations

Raman scattering activities were simulated adopting the Kohn-Sham formulation of the density functional theory (DFT)[35,36] as implemented in the Gaussian 03 program (version D.01)[37]. The input HAT6 geometry (144 atoms) was built by considering structures of benzene and hexalkoxy-groups for the aromatic core and tails (R=$OC_6H_{13}$), respectively. An initial fully planar $D_{3h}$ symmetry was assumed, the only out-of-plane atoms being the hydrogens of the tails. All calculations were performed on a Beowulf Intel cluster at the Delft University of Technology (the Netherlands). Several combinations of exchange-correlation (XC) functionals (local, semilocal and nonlocal) and basis sets (including or not polarization and diffuse functions) have been tested to establish the best model calculation(s) that lead to an explanation of experimental observations. The technicalities and methodologies used in the calculations are beyond the scope of the current topic and will not be discussed here. The predicted Raman activities were simulated in the gas phase adopting the SVWN XC functional, which consists of the Slater exchange (S)[38] combined to the Vosko, Wilk, and Nusair approximation (VWN) of the correlation part[39]. Pople's group basis set 6-311G** was adopted for all atomic types.[40] A Gaussian function of a FWHM of 10 $cm^{-1}$ was convoluted with the calculated Raman data to take account the resolution in the measured spectra.

### III. RESULTS

#### A. Absorption.

It is well established that discotic compounds change color upon complexation with an electron acceptor.[3,29] For HAT6-TNF this color change is quite strong: the 1:1 LC mixture of the white colored HAT6 and the yellow electron acceptor TNF becomes black. Indeed, the absorption spectrum of HAT6-TNF (Fig. 2 (a)) shows a broad excited CT-band extending from ~500 $nm$ to about 870 $nm$, the band gap thus being below 1.43 eV. In contrast, HAT6 shows a strong absorption band at 366 $nm$ and only weakly allowed transitions at longer wavelengths. Triphenylene and HAT6 absorption spectra have already been studied extensively in literature.[41-43] Based on the similarity with the present liquid crystalline measurement, we assigned the small absorption peaks at 469 and 442 $nm$ to the formally forbidden $S_0 \rightarrow S_1$ ($A'_1 \rightarrow A'_1$) electronic transition, a shoulder around 417 $nm$ to $S_0 \rightarrow S_2$ ($A'_1 \rightarrow A'_2$), the broad peak at 402 $nm$ to $S_3$ ($A'_1 \rightarrow E'$) and the absorption maximum at 366 $nm$ to the strong allowed $S_4$ ($A'_1 \rightarrow E'$) transition. The $S_4$ transition of HAT6 also dominates the higher energy region in the excited CT complex. The electronic absorption spectrum of TNF has been studied elsewhere,[44,45] and contains a weak lowest-energy band at a wavelength of 435 $nm$ due to the n-$\pi$* transition and higher energy bands at 387, 302, 260, and 222 $nm$ due to $\pi$-$\pi$* transitions. Thus, the 500-870 $nm$ region where the excited CT-band is situated is clear of any electronic transitions from pure HAT6 or TNF. The CT band is a result of the excited charge-transfer interaction between HAT6 and TNF involving electronic transition from the highest occupied molecular orbitals (HOMOs) of HAT6 to the lowest unoccupied molecular orbitals (LUMOs) of TNF, although the molecular orbitals become mixed in the excited CT complex.[31a,b]

#### B. Raman spectroscopy.

A complete overview of the off-resonance Raman spectra for HAT6, TNF and HAT6-TNF is presented in the Supplementary Material Document.[46] We assigned TNF by using DFT calculations in combination with analyses presented in earlier studies.[44,47,48] The most relevant Raman modes (Table S3) are due to symmetric C-$NO_2$ stretching (~1350 $cm^{-1}$), C-C skeleton vibration (1601 $cm^{-1}$) and C=O stretching (1733 $cm^{-1}$). There is good agreement between the measured and simulated Raman data (Fig. 2 (b)) for HAT6, which enabled us to assign most of the bands. Of prime importance is the E′ symmetric vibration IV at 1617 cm$^{-1}$, which is situated at the aromatic core of HAT6. It bears a strong resemblance to the quinoidal $\nu_8$ ($E_{2g}$) mode of benzene[24,25,49] as illustrated in Fig. 2 (d) and Fig. S2.

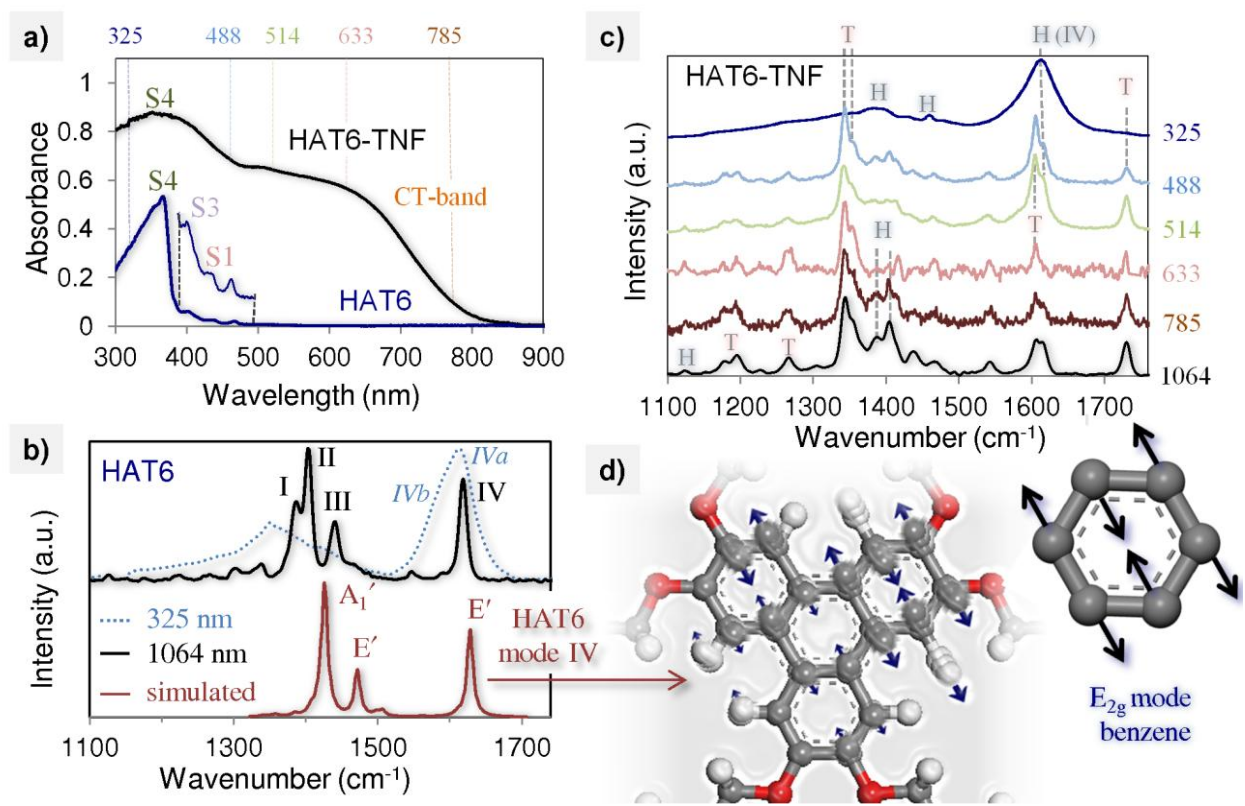

FIG. 2. Overview of the spectroscopic results. (a) Absorption spectra of HAT6 at 350 K and HAT6-TNF at 300K, including the assignment of the absorption bands. (b) Top: off-resonance (1064 *nm, solid line*) and Resonance (325 *nm, dotted line*) Raman spectra for HAT6 at 300 K. Bottom: off-resonance Raman activities profile calculated with DFT. (c) Raman spectra for HAT6-TNF at 300K for the indicated wavelengths. The assignment of the peaks is labeled as H for HAT6 and T for TNF. (d) Illustration of the E' normal mode calculated at 1628 cm$^{-1}$ (left) and its similarity with the quinoidal $E_{2g}$ mode of benzene (right).

The off-resonant Raman spectrum of the ground state CT-complex is a superposition of the HAT6 and TNF vibrational modes, which were assigned by comparison with the spectra of the uncomplexed components.[46] However, several vibrational modes are considerably shifted in frequency, and it is well established that these shifts of donor and acceptor frequencies are related to the occurrence of a ground state charge transfer.[48,50-54] In particular, for the acceptor, TNF, a redshift of the C=O stretching mode has been observed in ground state CT complexes, and is related to an increase in the electron density resulting from partial ground-state CT.[48,52] In the HAT6-TNF spectrum, the C=O mode is observed to have red-shifted to 1730 cm$^{-1}$. Based on the data in literature[52] it is estimated that the shift of 3 cm$^{-1}$ corresponds to an increase in electron density on TNF of about 0.03 e$^-$. In addition, for all the assigned modes we found satisfactory agreement between the observed shifts and normalized frequency changes between neutral TNF and the anion TNF$^{-1}$ calculated with DFT (Table S4), with the normalization factor corresponding to a ground state charge transfer of about 0.06 electron.

The electronic transitions in the excited CT complex and the vibrational modes involved in these transitions were investigated by exciting at different wavelengths in the absorption band (Fig. 2a). The corresponding resonant Raman spectra are shown in Fig. 2c (1100-1800 cm$^{-1}$ region) and Fig. S6 (500-1000 cm$^{-1}$).[46] The resonance Raman spectra at 633 and 785 *nm* were obtained after subtraction of a broad luminescence background (Fig. S5), which results in a lower signal to noise ratio. The resonance Raman spectra are dominated by the TNF vibrational bands, especially the symmetric C-NO$_2$ stretching modes around 1350 cm$^{-1}$ when exciting in the low energy region of the excited CT-band (785 and 633 *nm*). Most of the HAT6 vibrational modes (e.g. I, II, III and IV) are considerably less intense, or even absent, compared with the off-resonance case. However, the low-frequency radial breathing mode of the HAT6 aromatic core at 721 cm$^{-1}$ shows significant intensity for both excitation wavelengths (Fig. S6).[46] The activity of the symmetric C-NO$_2$ vibrations and the radial breathing mode of the HAT6 core suggest that the lowest electronic transition in the excited CT complex is due to a charge transfer from the HAT6 core to TNF, with a strong involvement of the nitro groups. Strong resonant activity of symmetric C-NO$_2$ stretching modes is also observed for small aromatic nitro compounds with intramolecular ground state CT.[55] In these molecules the π-π$^*$ electronic transition of the lowest excited state gives rise to significant bond length and bond angle changes in the C-NO$_2$ group, reflecting that the excited state wavefunction contains a large contribution from the basis functions of the nitro group.[55-59] For HAT6 it is well established that the HOMO is located on the aromatic core.[60,61] Based on these considerations, we propose that the lowest excited state in HAT6-TNF forming the lower energy region of the excited CT-band is predominantly a π-π$^*$ type transition involving the highest occupied molecular orbitals (HOMO, HOMO-1, etc.) of HAT6 and the lowest unoccupied molecular orbitals from TNF (LUMO, LUMO+1, etc),[31b] with a prominent role of the TNF nitro groups. Additional support for this assignment is obtained from DFT calculations on TNF,[46] showing that the LUMO is of $\pi^*$-type and contains a predominant contribution from the nitro groups.

The resonant spectra, after exciting higher energy in the CT-band, are significantly different from the spectra at 785 and 633 nm, although there is still significant activity of the symmetric C-NO$_2$ stretching modes for excitation at both 514 and 488 nm. But the most strongly enhanced TNF mode, compared to off-resonance, is the C-C skeleton vibration at 1604 cm$^{-1}$, whilst for HAT6 there is a strong resonant activity of the quinoidal vibration IV. The broad excited CT band thus seems to consist of a superposition of at least two different electronic transitions, involving different molecular orbitals of HAT6 and TNF. Such a superposition agrees with the small shoulder observed in the excited CT-band around 510 nm (Fig. 2 (a)) and partly explains the significant width of this band. However, from the similarity of the 785 and 633 nm spectra it also appears that the separate transitions give rise to rather broad absorption signals, which is consistent with the significant dynamic disorder in donor-acceptor distances found in the structural study.[32]

For excitation at 325 *nm* we can compare the CT-compound directly with the pure HAT6 and TNF compounds under resonance conditions. The resonant Raman spectrum of HAT6-TNF (Fig. 2 (c)) compares well with that of pure HAT6 (Fig. 2 (b)), both being significantly broadened and showing an envelope of bands with a maximum at about 1360 cm$^{-1}$ for HAT6 and 1380 cm$^{-1}$ for HAT6-TNF. The 325 *nm* spectrum of TNF, on the other hand, looks rather different with considerably narrower lines (Fig. S7).[46] These observations are all consistent with the proposition that the electronic transition relating to the high energy (~ 366 *nm*) absorption band of HAT6-TNF strongly resembles the $S_4$ ($A'_1 \rightarrow E'$) transition of pure HAT6. The most enhanced mode in the 325 *nm* spectra of HAT6 and the excited CT-complex is the quinoidal $E'$ symmetric HAT6 mode at about 1616 cm$^{-1}$. For pure HAT6 a shoulder is also clearly visible on this band, with a maximum considerably below 1600 cm$^{-1}$. The off-resonance spectrum of HAT6, on the other hand, only shows a weakly Raman-active mode at about 1592 cm$^{-1}$, in addition to mode IV at 1617 cm$^{-1}$. Similar observations exist for HAT6-TNF, although the shoulder is less clearly visible and a small contribution of the TNF C-C skeleton mode cannot be excluded. The 1500-1700 cm$^{-1}$ region of both spectra were fitted with two Voigt lineshapes; denoted with IVa and IVb (Fig. 3 and Table 1). An extra lineshape representing the TNF C-C skeletal mode was included for HAT6-TNF, with a fixed frequency 1605 cm$^{-1}$. The fitted peak positions of mode IVa are in good agreement with the observed frequencies for the $E'$ symmetric vibration in the off-resonance spectra of HAT6 (1617 cm$^{-1}$) and HAT6-TNF (1615 cm$^{-1}$). The fitted intensity for the TNF C-C skeletal mode is very small (~ 1% of the total area), confirming that the 325 *nm* resonant enhancement of TNF is not significant in the excited CT-complex (see also Fig. S7). Most importantly, the fit shows that a strong contribution of a second mode IVb must be present in both spectra, with a frequency of about 1585 cm$^{-1}$ (1584 cm$^{-1}$) for HAT6 (HAT6-TNF). The integrated intensity of the IVb is about 44 % of the total peak area for pure HAT6, and about 24% for HAT6-TNF.

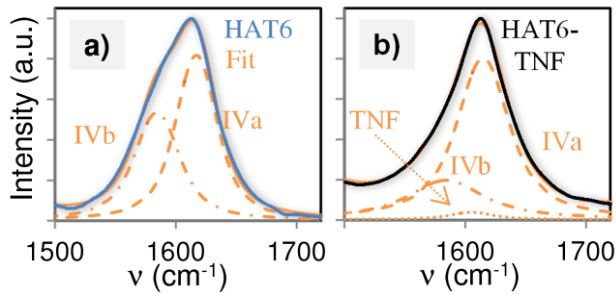

FIG. 3. Resonant Raman spectra of the E′ vibration IV for HAT6 (a) and HAT6-TNF (b) at 325 *nm*, fitted with Voigt lineshapes (orange). The mode is split into a IVa (dashed) and a vibronic IVb (dash-dotted) channel for both HAT6 and HAT6-TNF. The fitted contribution of the TNF band at 1606 cm$^{-1}$ (dotted line) is very small.

Table I. Results of the fit shown in Fig. 3

| Mode | Position (cm$^{-1}$) | FWHM (cm$^{-1}$) | Area (%) |
|---|---|---|---|
| IVa HAT6 | 1616.8 ± 0.5 | 41 ± 2 | 56 ± 5 |
| IVb HAT6 | 1585.1 ± 0.9 | 47 ± 4 | 44 ± 5 |
| IVa CT | 1615.1 ± 0.7 | 50 ± 5 | 75 ± 8 |
| IVb CT | 1584 ± 1 | 60 ± 9 | 24 ± 8 |

## IV. DISCUSSION

### A. Electron transfer in the ground state.

Intermolecular charge transfer in the electronic ground state is a well-known phenomenon for small π-conjugated molecules.[30,31] But in contrast, there are only a few reports on intermolecular ground-state CT in large molecular complexes, most of them involving polymers doped with an strong electron acceptor.[53,62] For discotic liquid crystalline compounds, involving D-A interaction, it is generally accepted that intermolecular charge transfer occurs in the excited state, but not in the ground state.[27,28] However, we have found strong indications for a weak ground-state electron transfer in the HAT6-TNF complex. Both the observed NMR chemical-shift changes reported in the previous article[32] and Raman frequency shifts are consistent with weak electron transfer from the HAT6 core to TNF, even leading to a comparable estimation for the amount of charge involved, which is about $6 \times 10^{-2}$ electron. To our knowledge, this is the first time that DFT calculations and two different experimental techniques have been used simultaneously to estimate ground-state CT effects. The strength of such combined analyses can be appreciated by considering the consistency on a more detailed level. The change in both Raman and NMR chemical shifts indicate that the electron transfer between HAT6 and TNF leads to a delocalized redistribution of the charge on TNF. For HAT6, the strongest changes in the NMR shifts and the Raman frequencies occur in the aromatic core. Furthermore, the TNF carbonyl vibration has a characteristic behavior: introduction of the electronegative $NO_2$ substituents into the fluorenone molecule tends to increase the C=O frequency, which can amount to ~25 cm$^{-1}$ within the series from aminofluorenone to TNF.[48,63] The observed Raman redshift of the C=O frequency in HAT6-TNF implies that the electron-withdrawing action of nitro substituents can be partly compensated by extra electron density donated by HAT6. This is further supported by the strong chemical shift change of the C=O carbon observed with NMR.

The ground state CT in combination with the polarity of the TNF molecules concurs with the observation of a large dipole moment for HAT6-TNF in dielectric relation spectroscopy measurements.[23] By using an estimated distance of 4 Å between HAT6 and TNF,[32] the dipole moment corresponding to a ground state transfer of $6 \times 10^{-2}$ electron is about 1.2 *D*. The presence of such permanent dipoles can be an important factor in facilitating charge separation of photo-generated excited CT states.[10,13] It has been shown that a dipolar layer at the D-A interface can lead to a repulsive barrier separating the hole and electron residing on neighboring donor and acceptor molecules.[14]

### B. Vibrational relaxation processes in the excited states

In section III we used the resonant activity of specific molecular vibrations in the Raman spectra (Fig. 2) to characterize the different electronic excited states of HAT6-TNF. Further, the resonant Raman spectra offered an additional opportunity to identify some of the charge carrier relaxation processes in the excited states along with the related molecular vibrations involved in these processes. First, we consider the significantly broadened spectra obtained in the resonance mode after excitation at 325 *nm*, with the $S_4$ ($A'_1 \rightarrow E'$) electronic transition of HAT6. For both HAT6 and HAT6-TNF we found a strong enhancement of the *e'* symmetric quinoidal mode IV, accompanied by the appearance of a second band IVb at 1585 cm$^{-1}$. Such an activity of non-totally symmetric *e'* modes is notable, and has been observed already for the smaller building blocks of HAT6, triphenylene and benzene. Both triphenylene and benzene are so called (*A+E*) × *e* systems, in which a doubly degenerate electronic state *E* is vibronically coupled (pseudo-Jahn-Teller) to nondegenerate states *A* through the degenerate *e* vibrational modes.[64,65] In the case of triphenylene, the *e'* mode around 1600 cm$^{-1}$ has been shown to be the main channel facilitating pseudo-Jahn-Teller interactions between the lowest lying triplet states.[24,49,64] In addition, it has been found that this Jahn-Teller mode provides a significant contribution to the reorganization energy of HATn molecules, which is a limiting factor for hole transport along the columnar stacks.[61] For benzene it is well established that the $\nu_8$ ($e_{2g}$) mode couples not only the lowest triplet states, but also the lowest singlet states $B_{2u}$ and $B_{1u}$ with the doubly degenerate $E_{1u}$ state.[49,65] Moreover, the $\nu_8$ mode splits into two channels $\nu_{8a}$ and $\nu_{8b}$ in resonance with the strongly allowed ($A_{1g} \rightarrow E_{1u}$) transition when there is an OH or O-substituent attached to the benzene ring.[66] The presence of the $\nu_{8b}$ component implies vibronic activity[65,66] since this channel gains intensity entirely from Albrecht's B-term[67]. Based on the strong parentage of the HAT6 *e'* mode IV with the quinoidal $\nu_8$ benzene vibration, we propose that a similar vibronic mechanism is responsible for the enhancement of two components IVa and IVb, in resonance, with the $S_4$ ($A'_1 \rightarrow E'$) band. We have found that mode IV of HAT6 is dominated by the $\nu_8$ type of motion on the outer aromatic rings (Fig. 2), resembling the above situation of benzene with an O or OH substituent. Furthermore, the positions of the two channels IVa (~1616 cm$^{-1}$) and IVb (1585 cm$^{-1}$) are consistent with the relative frequencies of the $\nu_{8a}$ and $\nu_{8b}$ modes observed for the substituted benzenes, being in the range of 1589-1617 cm$^{-1}$ ($\nu_{8a}$) and 1560-1601 cm$^{-1}$ ($\nu_{8b}$) with a mutual separation of 15-30 cm$^{-1}$. As with the lower frequency mode $\nu_{8b}$ in benzene, the IVb channel must gain its intensity entirely from vibronic (Albrecht's B-term) activity. The presence of this mode for excitation at 325 *nm* is therefore a strong indication that the quinoidal motion of the aromatic core is the main channel for intramolecular relaxation through vibronic coupling with lower lying electronic states. The strong broadening of the 325 *nm* spectra of HAT6 and HAT6-TNF indicates that the electron-phonon coupling processes that are involved, are fast on the Raman timescale. This is supported by the considerable Lorentzian contribution needed to obtain a good fit of the resonant spectrum, indicating a homogeneous line broadening effect.

The Raman linewidths may be used to estimate the timescale, $\tau$, of the relaxation processes within the excited state, via the energy-time uncertainty relation $\tau = \hbar/\Delta E$. $\Delta E$ being the linewidth in cm$^{-1}$ and $\hbar = 5.3\times10^{-12}$ cm$^{-1}$ s.[68] By taking the fitted FWHM of the IVb peaks (Table 1) we obtain a timescale of 113 fs for HAT6 and 86 fs for HAT6-TNF. It seems that the excited-state relaxation processes are somewhat faster in the excited CT-complex. However, these are estimates of the minimum timescales, since other processes, like pure dephasing due to quasi-elastic events, may also contribute to the linebroadening.[69] For benzene it has also been observed that the relaxation processes within the $E_{1u}$ band are fast, typically with a decay rate of $10^{14}$ s$^{-1}$,[70,71], which is comparable to our lower bounds for internal conversion in the $E'$ band of HAT6.

Considering the excited CT-band, the enhanced ground-state vibrational modes in the resonant Raman spectra cannot be related directly to relaxation processes, as for the 325 *nm* spectra. However, we found that the symmetric NO$_2$ stretching vibration of TNF is the most prominent Raman active mode in the whole excited CT-band involving different electronic transitions, and that the lowest excited state involves charge transfer from the HAT6 aromatic core to TNF. From other studies it is apparent that considerable bond length and bond angle changes in the C-NO$_2$ group are involved in charge-transfer excitation of small aromatic nitro compounds[55,59]. It is therefore reasonable to expect that the localized vibrations of the nitro groups play a significant role in relaxation processes within the excited CT-band. Furthermore, it appears that the relaxation processes in the CT-band are much slower than in the higher energy band. Considering the Raman linewidths, we estimate that the relaxation processes in the CT-band must be at least on the picosecond timescale, considerably slower than for the high energy HAT6 band. We tentatively argue that hot-carrier relaxation processes in the CT-band in the visible light region are relatively slow compared to the fast relaxation within the original UV absorption band of pure HAT6, which can be relevant concerning the efficient separation of charges in organic PV-devices.

## V. CONCLUSIONS

We have found conclusive evidence for ground-state electron transfer in the prototypical discotic complex HAT6-TNF. The results from NMR and Raman were both consistent with weak electron transfer from the HAT6 core to TNF in the ground state, even leading to a comparable estimation for the amount of the charge involved, which is of the order of 6x10$^{-2}$ electron. It was shown that the excited CT-band of HAT6-TNF consists of different intermolecular electronic transitions. The lowest excited state was deduced to be predominantly a $\pi$-$\pi^*$ type of transition from the HAT6 HOMO on the aromatic core to the LUMO of TNF, the latter containing a significant contribution from the basis functions of the nitro groups. A high energy shoulder at 366 *nm* in the absorption spectrum of HAT6-TNF was observed and assigned to the strongly allowed $S_4$ (A$'_1$→E$'$) transition of pure HAT6.

We have identified a fast intramolecular relaxation process within this 'hot' $S_4$ band in both pure HAT6 and HAT6-TNF. This relaxation involves the quinoidal motion of the aromatic core, in close analogy with vibronic coupling mechanisms occurring in the building block benzene. The strong resemblance of the quinoidal relaxation process in the hot band of HAT6 to the case of benzene suggests that the underlying vibronic coupling mechanism is a fundamental aspect of polyaromatic hydrocarbons. In contrast, charge-carrier relaxation processes within the broad excited CT-band seem to be relatively slower than the fast internal conversion in the high energy intramolecular band of HAT6. Both the presence of permanent CT dipoles and slower relaxation processes in the CT band can be favorable concerning efficient charge separation in organic PV-devices.

## ACKNOWLEDGEMENTS

We gratefully acknowledge Lauren Clements and Elise Talgorn for support with the Raman and absorption measurements, and prof. Robert Armstrong for critical reading. This work is part of the research program of the Foundation for Fundamental Research on Matter (FOM), which is financially supported by the Netherlands Organization for Scientific Research (NWO). This article is the result of joint research in the Delft Research Centre for Sustainable Energy and the 3TU Centre for Sustainable Energy Technologies. The Renishaw inVia Reflex Raman spectrometer and the Bruker MultiRAM Raman spectrometer, located at The University of Sydney, were purchased using a LIEF grant (LE0560680 and LE0883036) from the Australian Research Council (ARC).

# Supporting information for:

# Electronic and vibronic properties of a discotic liquid-crystal and its charge transfer complex

Lucas A. Haverkate[1], Mohamed Zbiri[2,a)], Mark R. Johnson [2], E. A. Carter [3], Arek Kotlewski[4], Stephen J. Picken[4], Fokko M. Mulder[1], Gordon J. Kearley [5]

[1]*Reactor Institute Delft, Faculty of Applied Sciences, Delft University of Technology, Mekelweg 15, 2629JB Delft, The Netherlands.*

[2]*Institut Laue Langevin, 38042 Grenoble Cedex 9, France.*

[3]*Vibrational Spectroscopy Facility, School of Chemistry, The University of Sydney NSW 2006, Australia*

[4]*ChemE-NSM, Faculty of Chemistry, Delft University of Technology, 2628BL/136 Delft, The Netherlands.*

[5]*Bragg institute, Australian Nuclear Science and Technology Organisation, Menai, NSW 2234, Australia.*

a)*zbiri@ill.fr*

## Assignment of the Raman modes

**HAT6.** The measured and simulated off-resonance Raman spectra for HAT6 (protonated) and HAT6D (deuterated) are compared in Figure S1 and Table S1. There is good agreement between simulation and experiment, the prediction of the relative peak positions and intensities being very satisfactory. Also, the direction of the frequency shift of the normal modes due to deuteration is reproducible, although the magnitude of the shifts is overestimated for some of the peaks. The overall agreement allows us to assign the most important Raman modes as indicated in Table S1. The most important vibrational mode for our analyses is mode IV. Figure S2 shows that this normal mode is dominated by C-C and C-H vibrations of the triphenylene core, this being experimentally supported by the limited frequency shift upon deuteration. For triphenylene it is well established that the E′ mode around 1600 cm$^{-1}$ bears a strong parentage with the $\nu_8$ mode of benzene.[1,8,10]

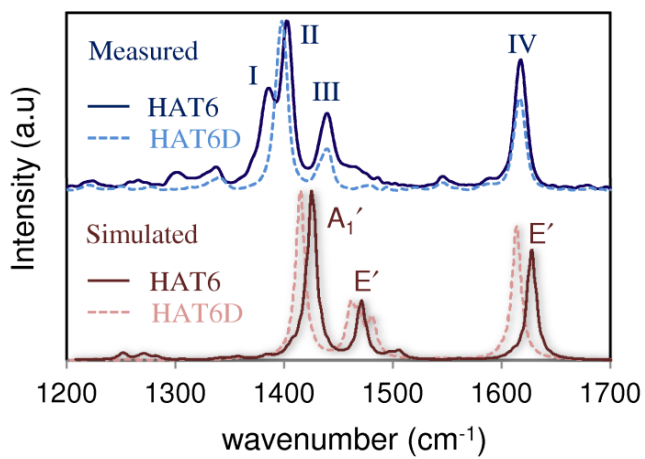

Figure S1 Measured (1064 nm, 300K) Raman spectra and simulated Raman activities for HAT6 (solid lines) and HAT6D (dashed lines).

Table S1 Simulated and observed HAT6 peak positions in *cm$^{-1}$*

|  | I | II | III | IV |
|---|---|---|---|---|
| Irreps (D$_{3H}$) | ? | A$_1'$ | E$'$ | E$'$ |
| HAT6 meas | 1386 | 1403 | 1439 | 1617 |
| HAT6D meas | - | 1398 | 1438 | 1616 |
| HAT6 sim | - | 1425 | 1471 | 1627 |
| HAT6D sim | - | 1415 | 1461 | 1614 |

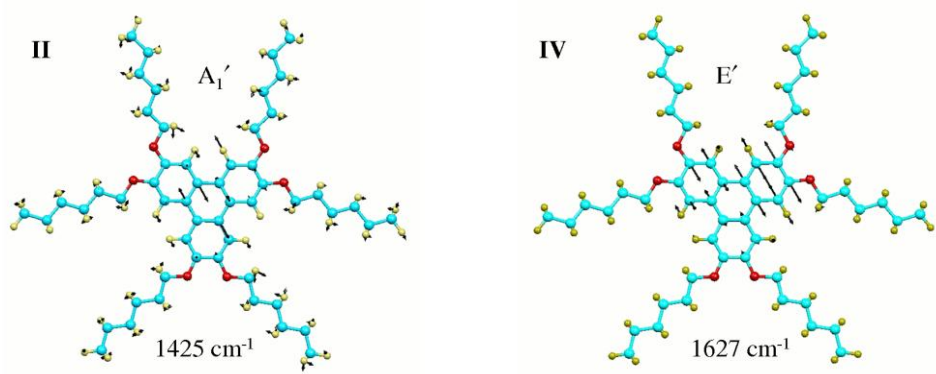

Figure S2 Illustration of the HAT6 II and IV modes

For HAT6 the simulated E′ mode at 1627 cm$^{-1}$ contains the characteristic quinoidal vibration of the benzene rings. The motion is not uniformly distributed over the rings, as the strongest displacements are located at the upper left ring in Figure S2. Peak II is assigned to A$_1$′ symmetric normal mode as illustrated in Figure S2. As well as the prominent core motion, this mode also contains a significant contribution from the tail hydrogens. This is reflected by the frequency shift upon deuteration found in both the experimental and simulation results. The normal mode III (E′ symmetry) bears a contribution from the tail hydrogen motion. For both mode II and III it is difficult to trace the aromatic ring motions back to fundamental modes of a single benzene molecule. Rather, these modes reflect the collective motion of the four aromatic rings, resulting in a deformation type of vibration.[11] Mode I at 1384 cm$^{-1}$ in the protonated sample could not be resolved in the simulation. There is a mode at 721 cm$^{-1}$ (720 cm$^{-1}$ for HAT6D) with significant intensity (Figure S4), which is assigned as the totally symmetric radial breathing mode of the aromatic core[5]. The C-H stretching modes of the core (3100-3200 cm$^{-1}$) and tails (2900-3000 cm$^{-1}$) are situated in the higher frequency region.

**HAT6-TNF.** Figure S3 shows the 1064 nm off-resonance Raman spectra of HAT6, TNF and HAT6-TNF recorded at 300K. The main vibrations in pure TNF have been studied elsewhere.[9,12,16], the group of three peaks around 1350 cm$^{-1}$ being assigned to symmetric stretching of the three NO$_2$ side groups of TNF. The peak at 1601 cm$^{-1}$ is the main C-C skeleton vibration of the TNF aromatic rings, whilst the peak at 1733 cm$^{-1}$ is due to the C=O stretching vibration. All the HAT6-TNF composite peaks, indicated with an arrow in Figure S3, were assigned to either HAT6 or TNF with the help of the deuterated analogs of the samples as shown in Figure S4.

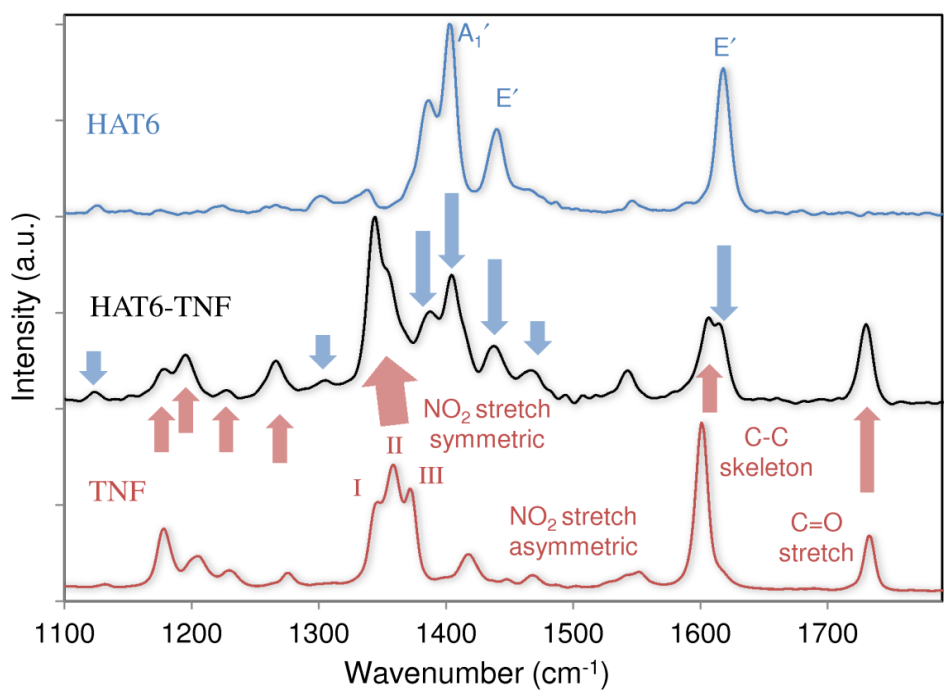

Figure S3 Off resonance Raman spectra recorded at 1604 nm for fully protonanted HAT6 (blue), TNF (red) and HAT6-TNF (black) at 300K.

Tables S2 and S3 compare the frequencies of the pure samples HAT6 and TNF with the fully protonated composite. The frequency shift of the TNF C=O stretching mode is 3 +/- 1 $cm^{-1}$. It was difficult to assign the $NO_2$ symmetric stretching peaks in the composite uniquely because their frequency shifts due to complex formation are comparable to their mutual separation, and the intensity distribution has been changed in the complex. The frequency assignment in Table S3 leaves the order of the peaks the same, giving rise to the smallest frequency shifts. Smaller peaks in the composite, such as the asymmetric $NO_2$ stretching group, were difficult to assign to the composite, but the aim of the present research is to assign the most prominent features. It is important for the resonant Raman analyses that there is no doubt about the two *off*-resonance peaks around 1600 $cm^{-1}$ in the composite: the left one is the TNF skeleton mode, the right one the E′ quinoidal vibration of the HAT6 benzene rings.

Table S2 Comparison of observed HAT6 peak positions in $cm^{-1}$, with an error of 1 $cm^{-1}$

|  | I | II | III | IV |
| --- | --- | --- | --- | --- |
| Irreps ($D_{3H}$) | ? | **A₁′** | **E′** | **E′** |
| HAT6 | 1386 | 1403 | 1439 | 1617 |
| HAT6-TNF | 1387 | 1404 | 1437 | 1615 |

Table S3 Comparison of observed TNF peak positions in $cm^{-1}$, with an error of 1 $cm^{-1}$

|  | NO$_2$ stretch symm I | NO$_2$ stretch symm II | NO$_2$ stretch symm III | C-C skeleton | C=O stretch |
|---|---|---|---|---|---|
| TNF | 1345 | 1358 | 1372 | 1601 | 1733 |
| HAT6-TNF | 1343 | 1354 | 1367 | 1605 | 1730 |

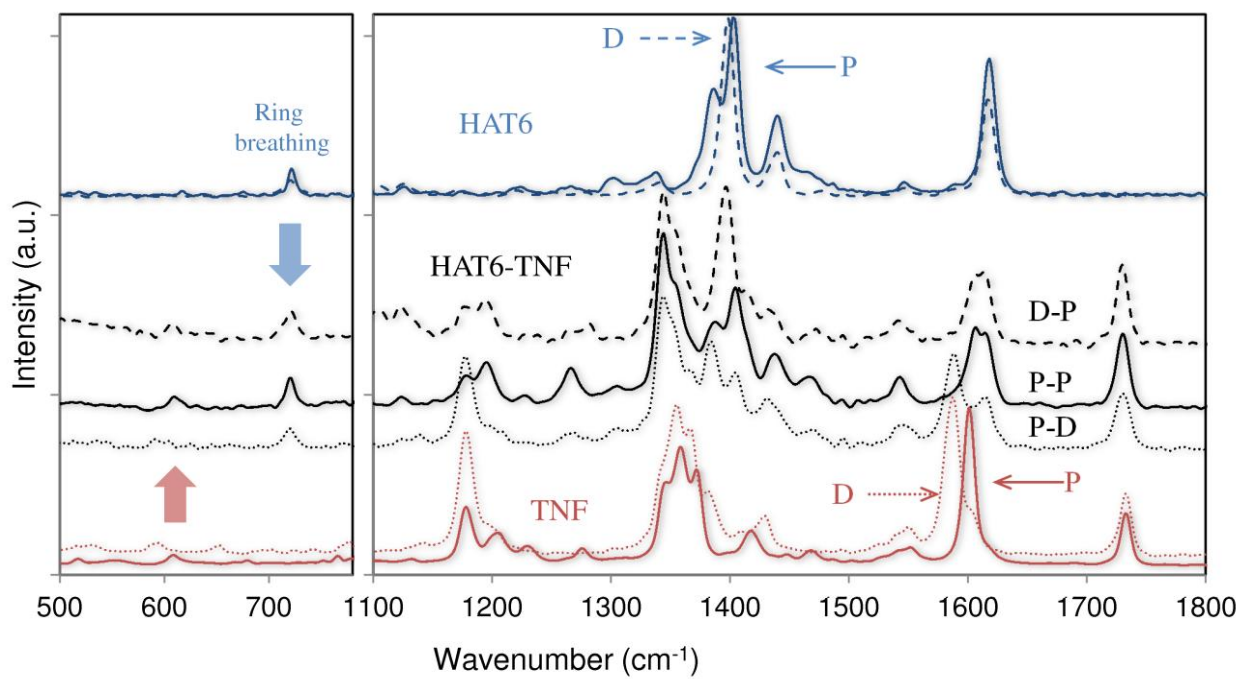

Figure S4 Off resonance Raman spectra recorded at 1604 nm for HAT6 (blue), HAT6D (blue dashed), TNF (red), TNFd (red dotted), HAT6-TNF (black), HAT6D-TNF (black dashed) and HAT6-TNFd (black dotted) at 300K. The band at 719 $cm^{-1}$ in the CT-complex spectrum corresponds to the symmetric aromatic ring breathing mode of HAT6 (721 $cm^{-1}$) as indicated with the blue arrow.

**Resonance Raman**

**Luminescence in HAT6-TNF.** The HAT6-TNF resonance spectra collected using excitation wavelengths of 633 and 785 *nm* in the article (Figure 7c) were obtained after subtracting a prominent luminescence background. Figure S5 presents the raw spectrum including the luminescence peak of HAT6-TNF collected at 785 nm. The peak maximum corresponds to an emission wavelength of 882 +/- 10 nm (1.41 eV), indicating that a large proportion of the incident photons decays with $\Delta E$ = 0.17 eV non-radiatively to the band edge. Note that the absorption band edge of the CT-band was observed (Figure 2a) at about 870 nm (1.43 eV). For 633 *nm* a similar luminescence (i.e. comparable width and relative intensity) background was observed, with the peak maximum towards higher wavenumbers as expected. However, the corresponding emission maximum at 810 +/- 10 *cm$^{-1}$* (1.53 eV) was significantly higher in energy than for excitation at 785 *nm*. It thus seems that the broad CT-band covers more than one electronic transition, with the shift of the emission maximum for lower excitation wavelengths being due to hot-luminiscence.

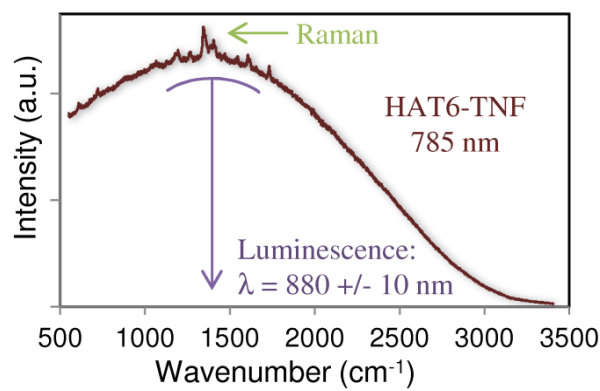

Figure S5 HAT6-TNF resonance Raman spectra recorded at 785 nm showing the broad luminescence background with an emission wavelength of 880 +/- 10 *nm*.

**HAT6-TNF Resonance Raman in the 550-1000 cm$^{-1}$ region.** Figure S6 presents the 550-1000 cm$^{-1}$ spectral region of the resonance Raman spectra of HAT6-TNF, the 1200-1800 cm$^{-1}$ region is presented in Figure 2c in the main article. The spectra are all normalized on the 1200-1800 cm$^{-1}$ region in a similar manner to Figure 2c. It is clearly that the HAT6 aromatic breathing mode at 719 cm$^{-1}$ is significantly enhanced when using excitation wavelengths of 785 and 633 nm. At shorter excitation wavelengths this mode is negligibly enhanced compared to the modes in the 1200-1800 cm$^{-1}$ region.

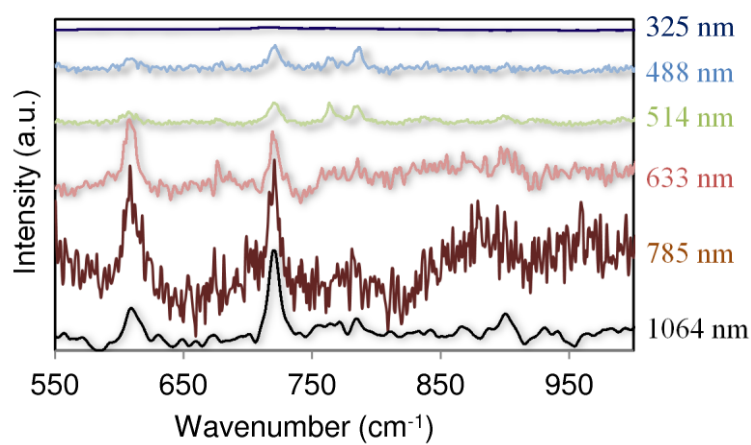

Figure S6 HAT6-TNF resonance Raman spectra for the selected wavelengths at 300K. The same spectra as shown in Figure Xc, but for the 550-1000 cm$^{-1}$ region.

**TNF at 325 nm.** Figure S7 shows the 325 *nm* resonant and 1064 *nm* off-resonant Raman spectra of TNF and HAT6-TNF. The main vibrational mode that is enhanced in pure TNF is the skeleton C-C motion at 1605 cm$^{-1}$. However, in the composite the intensities of TNF vibrations appear to be negligible. The contribution of the 1605 cm$^{-1}$ mode to the total integrated area of the spectral region around 1600 cm$^{-1}$ in HAT6-TNF is less than 1% (Figure 3 main article). Furthermore, the linewidths for pure TNF at 325 *nm* (about 20 cm$^{-1}$) are significantly smaller than for HAT6 and in the composite (~50 cm$^{-1}$ or even more).

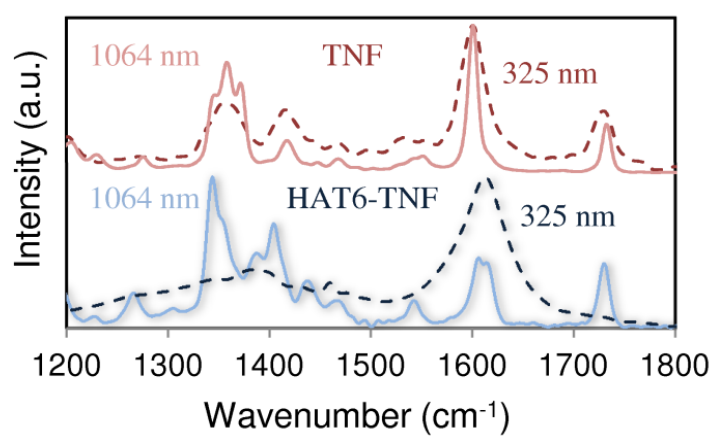

Figure S7 Off-resonance (solid lines) and 325 nm (dashed lines) Raman spectra of TNF (top) and HAT6-TNF (bottom)

**DFT on TNF**

To further support the assignment of the TNF electronic states and vibrational modes, we performed DFT calculations on neutral TNF and on the anion TNF$^{-1}$ with one extra valence electron. The DFT calculations were performed with the DMol$^3$ package as implemented in Materials Studio, using the PBE XC functional.[13] We checked that another XC functional (PW91[14]) gives comparable results. The DNP basis set[4] was adopted, the core electrons being described with Effective Core Potentials (ECPs).[2,6] The molecular configurations of TNF and TNF$^{-1}$ were obtained by ionic relaxation of the crystallographic structure of TNF[9], using the DFT settings described above. Subsequently, to obtain the molecular orbitals and vibrational modes, an energy minimization was performed with an SCF density convergence of 10$^{-6}$ and orbital cutoff of 3.7 Å. All calculations were performed spin-unrestricted, and in the gas phase.

Figure S8 shows the contour plot of the calculated HOMO and LUMOs of TNF. In line with the experimental observations discussed in this work and in literature,[7,9] the HOMO and LUMO correspond to a lowest electronic transition from the carbonyl oxygen non-bonding orbitals to a delocalized π* orbital. It is also observed that the LUMO and LUMO+1 contain a predominant contribution from the basis functions of the C-NO$_2$ groups. The LUMO-HOMO energy difference for TNF was calculated at 2.2 eV, which is slightly too small in comparison with the observed lowest absorption band of TNF at 2.85 eV. However, it is well established that GGA XC functionals generally underestimate the band gap,[3] and more sophisticated DFT approaches will certainly lead to a better quantitative agreement. Finally, we note that, as expected, the calculated HOMO of the anion TNF$^{-1}$ is very similar to the LUMO of TNF.

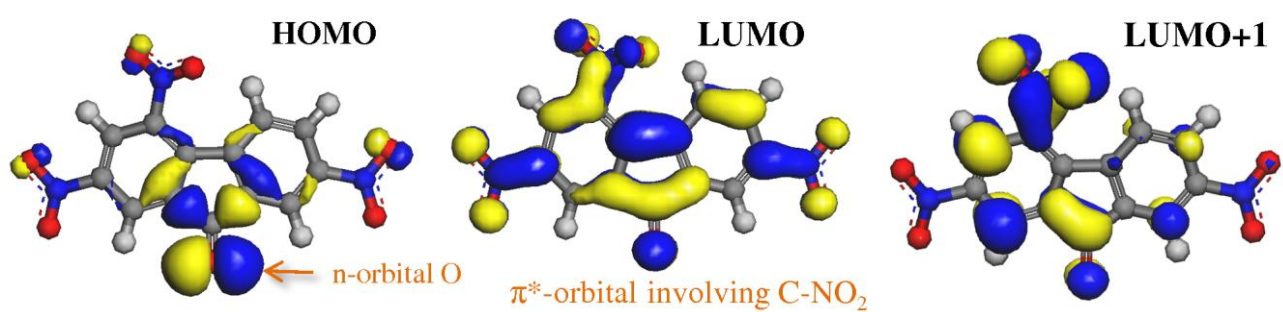

Figure S8 Illustration of the HOMO and LUMOs of TNF calculated with DFT.

Figure S9 and S10 illustrate the calculated TNF normal modes relevant for the above Raman analysis, i.e. the symmetric C-NO$_2$ stretching, C-C skeleton and C=O stretch vibrations. Overall, there is a satisfactory agreement between the calculated frequencies and the values observed with Raman spectroscopy (Table S3), and the results are consistent with the assignment made in Figure S3 which was based on literature.[9,12,16] Figure S9 shows that the C-NO$_2$ stretching modes are localized on the nitro groups. In addition to the symmetric NO$_2$ stretching, these normal modes also include stretching of the C-N bond and a small bending of the NO$_2$ group. The C-C skeleton vibration calculated at 1602 cm$^{-1}$ (Figure S10) contains a quinoidal type of vibration of the benzene rings, which explains why the frequency of this mode is close to the observed and calculated value of the quinoidal HAT6 mode IV. It is noted that this normal mode, although dominated by the displacements of the carbons, also contains small displacements of the hydrogens and NO$_2$ groups.

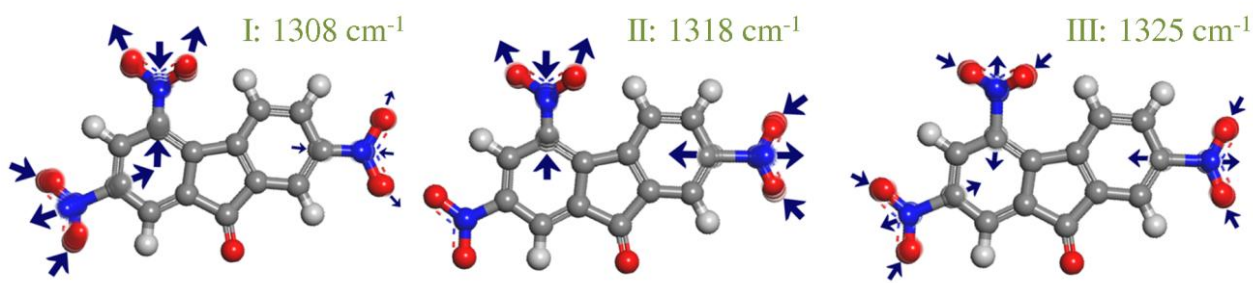

Figure S9 Illustration of the C-NO$_2$ stretching modes of TNF, calculated with DFT.

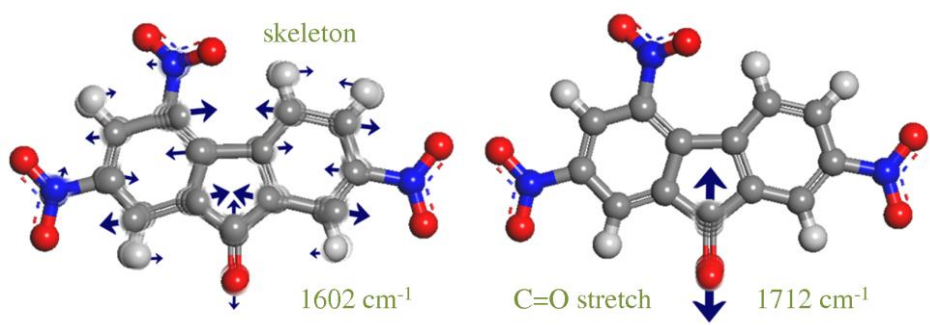

Figure S10 Illustration of the skeleton and C=O stretch modes of TNF, calculated with DFT.

Table S4 Calculated vibrational frequencies for TNF and TNF$^{-1}$, the difference $\Delta$ between these values, normalization of this difference w.r.t. the observed CT shift of the C=O frequency for HAT6-TNF (0.06×$\Delta$), and the experimentally observed CT shifts in HAT6-TNF (c.f. Table S3). All values are given in cm$^{-1}$.

|  | NO$_2$ stretch symm I | NO$_2$ stretch symm II | NO$_2$ stretch symm III | C-C skeleton | C=O stretch |
|---|---|---|---|---|---|
| TNF (=$v_T$) | 1308 | 1318 | 1325 | 1602 | 1712 |
| TNF$^{-1}$ (=$v_{T^{-1}}$) | 1229 | 1260 | 1275 | 1613 | 1662 |
| $\Delta$ (=$v_{T^{-1}}$ - $v_T$) | -79 | -58 | -50 | 11 | -51 |
| 0.06×$\Delta$ | -5 | -3 | -3 | 1 | -3 |
| CT shifts (exp.) | -2 | -4 | -5 | 4 | -3 |

We also calculated the vibrational modes for the case that one additional electron is placed on TNF, i.e. for the cation TNF$^{-1}$. Table S4 compares the calculated frequencies of the relevant vibrations for TNF$^{-1}$ with neutral TNF. The additional electron on TNF results in significant frequency shifts ($\Delta$) of the vibrational modes. Most importantly, the direction of the calculated shifts (redshift of the C-$NO_2$ and C=O stretching frequencies, blueshift of the C-C skeleton frequency) are similar to the observed Raman shift directions for the TNF modes in the CT-complex HAT6-TNF. By normalizing the calculated shifts Δ for the anion TNF$^{-1}$ to the observed CT-shift of the C=O stretching vibration, a multiplication factor of 0.06 is obtained. As can be seen in Table S4, all the normalized shifts (0.06×Δ) are in reasonable agreement with the frequency shifts observed for HAT6-TNF. Based on the DFT calculations it is thus estimated that the observed CT-shifts for HAT6-TNF correspond to a ground state charge transfer of about 0.06 electron. As a first approximation, this value compares well with the estimation of 0.03 electron based on data in literature[15]. Therefore, the DFT calculations on TNF strongly support the assignment of the observed Raman shifts for HAT6-TNF to ground state charge transfer.